\documentclass[twocolumn,superscriptaddress,notitlepage]{revtex4-1}
\usepackage{amsmath}
\usepackage{amssymb}
\usepackage{bm}
\usepackage{epsfig}
\usepackage{graphicx}
\usepackage[colorlinks]{hyperref}

\newcommand{\heading}[1]{\paragraph*{#1}}

\begin{document}     

\title{Spin dynamics of hopping electrons in quantum wires: algebraic decay and noise}

\author{A.V. Shumilin}
\affiliation{Ioffe Institute, 194021 St.-Petersburg, Russia}

\author{E.Ya. Sherman}
\affiliation{Department of Physical Chemistry, The University of the Basque Country, 48080 Bilbao, Spain}
\affiliation{IKERBASQUE Basque Foundation for Science, Bilbao, Spain}

\author{M.M. Glazov}
\affiliation{Ioffe Institute, 194021 St.-Petersburg, Russia}
\affiliation{Spin Optics Laboratory, St.-Petersburg State University, 1 Ul'anovskaya, 198504 Peterhof, St.-Petersburg, Russia}

\begin{abstract}
We study theoretically spin decoherence and intrinsic spin noise in semiconductor quantum wires caused by an interplay of electron hopping between the localized states and the hyperfine interaction of electron and nuclear spins. At a sufficiently low density of localization sites the hopping rates have an exponentially broad distribution. It allows the description of the spin dynamics in terms of closely-situated ``pairs'' of sites and single ``reaching'' states, from which the series of hops result in the electron localized inside a ``pair''. The developed analytical model and numerical simulations demonstrate disorder-dependent algebraic tails in the spin decay and power-law singularity-like features in the low-frequency part of the spin noise spectrum. 
\end{abstract}

\date{\today}
\maketitle

\heading{Introduction.} Recent progress in semiconductor nanotechnology and request in new hardware elements 
for prospective quantum technology devices caused a strong interest in one-dimensional solid-state systems such as 
semiconductor nanowires and nanowire-based 
heterostuctures. There are at least two main reasons for this interest. First reason is related to the abilities to produce and controllably 
manipulate electron spin qubits in InSb and InAs nanowires \cite{nadjperge2010,vanderberg2012} and 
superconductor-semiconductor-nanowire hybrids \cite{larsen2015}. 
Second reason is that, InSb nanowires, due to a strong spin-orbit coupling, are important 
as the hosts of edge Majorana states in such hybrid structures \cite{oreg2010,das2012,brouwer2011,degottardi2013}. 
These states, being of a fundamental interest for the quantum theory, 
are thought to be promising for the quantum computation applications as well. 
Disorder in the nanowires can play a crucial role in the physics of the 
qubits and Majorana states~\cite{adagideli2014}, and a variety of disorder regimes, dependent on the growth procedure and doping, 
is possible \cite{pitanti2011}. On the other hand, one-dimensionality leads to a strong localization of carriers, and the understanding 
of properties of localized electron states is requested for the understanding of these systems. 
In addition to the spin-orbit coupling, which can be reduced by choosing an appropriate
nanowire realization and geometry, III-V semiconductors show a strong hyperfine interaction, that is a source 
of the spin dephasing of localized electrons \cite{merkulov02,PhysRevLett.88.186802}. At a finite temperature, 
electrons can hop between different sites, and, 
therefore, experience randomly fluctuating hyperfine fields. 
Therefore, the hopping leads to spin relaxation and, thus, results 
in a spin noise. Recently, spin-noise spectroscopy experiments~\cite{aleksandrov81,zapasskii13,oest:rev} 
have made it possible to access information about new 
regimes of spin dynamics, including a very long time evolution of electrons 
and nuclei~\cite{crooker2010,PhysRevLett.115.207401,2015arXiv150605370B}. 
The width of the spin-noise spectrum is related to the relaxation rate and the spectrum features seen 
as deviations from the Lorentzian shape can reveal various non-Markovian 
memory effects~\cite{PhysRevLett.107.156602}. 

Here we address the spin relaxation and spin noise 
in semiconductor quantum wires due to the random hyperfine coupling induced by electron hopping between localized states. 
This disorder-determined relaxation is long and strongly non-exponential. The understanding of this spin relaxation 
mechanism can be valuable for the analysis of charge and spin transport 
in semiconductor nanowires and related hybrid structures.

\begin{figure}
\includegraphics[width=\linewidth]{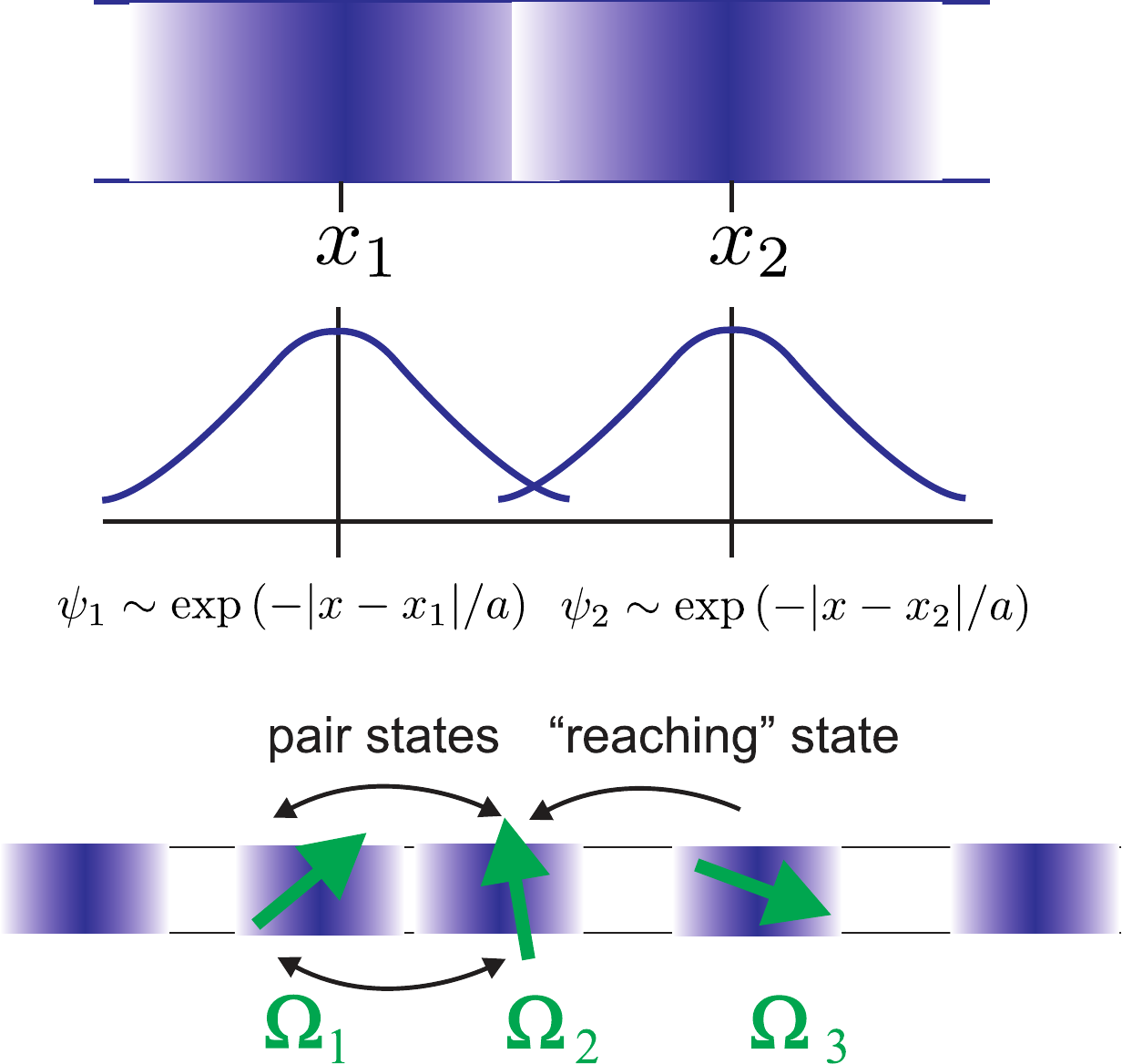}
\caption{Upper part: Sketch of a nanowire with localized states, schematically presenting the wavefunctions beneath 
the nanowire. Lower part: Illustration of a ``pair'' and a ``reaching'' 
state with hyperfine fields ${\bm\Omega}_{i}$ (green arrows).}\label{fig:model}
\end{figure}

\heading{Model.} We consider a nanowire where the electrons are assumed to be localized at single donors 
or fluctuations of the wire width. The density of localization sites $n$ and the 
localization length $a$ satisfy the condition
\begin{equation}
\label{kinetics}
na \ll 1,
\end{equation}
meaning a weak overlap of the wavefunctions, Fig.~\ref{fig:model}. 
Electrons hop between the sites with the aid of acoustic phonons. 
At each localization site $i$ the electron 
spin experiences a random static effective magnetic field with the precession 
frequency $\bm \Omega_i$. These frequencies are uncorrelated and isotropically 
distributed, $\langle \Omega_{i,\alpha} \Omega_{j,\beta} \rangle = \delta_{ij} \delta_{\alpha\beta} \Omega_0^2/3$, 
where $\alpha,\beta = x,y,z$ enumerate Cartesian components, the distribution of precession frequencies at a given site is 
Gaussian with the root-mean-square of~$\Omega_0$~\cite{merkulov02,gis2014noise,gi2012noise}.
Under the condition~\eqref{kinetics} the electron spin density matrix can be parametrized by the 
average occupancies, $N_i$, and spins, $\bm S_i$, at the sites.
Here we consider small electron concentration, $N_i \ll 1,$ neglect
electron-electron interactions, and assume that each site is either empty 
or singly occupied~\cite{suppl}. The spin dynamics 
can be described by the set of kinetic 
equations~\cite{PhysRevB.92.014206,PhysRevB.91.195301,suppl}: 
\begin{equation}
\label{set}
\frac{d \bm S_i}{dt} + \bm S_i \times \bm \Omega_i = \sum_j W_{ij}(\bm S_j - \bm S_i),
\end{equation} 
with $W_{ij}$ being the hopping rates between the sites $i$ and $j$. The spin-orbit interaction 
and hyperfine-irrelated spin-flip processes are disregarded. 
For the hopping rates we take the minimal model by assuming
\begin{equation}\label{Wij}
W_{ij} = \frac{1}{\tau} \exp\left(-2\frac{|x_i-x_j|}{a}\right),
\end{equation}
where $x_i$ and $x_j$ are the coordinates of the localization sites $i$ and $j$, 
respectively, and $\tau$ is a prefactor governed by the strength of the 
electron-phonon interaction. The spread of $a$ and $\tau$ is disregarded here~\cite{suppl}. 
Kinetic equation~\eqref{set} enables us to evaluate the 
spin dynamics and spin fluctuations in the nanowire.

The condition~\eqref{kinetics} leads to an exponentially broad spread of the hopping rates. 
This effect together with possible multiple returns of hopping electrons to
their initial sites determines the major specific 
features of the spin dynamics in nanowires. To analyze these 
phenomena in detail we further consider a realistic situation of rare transitions between the 
sites, $\Omega_0\tau \gg 1$, assuming, however, that the nuclear spin dynamics controlled by the hyperfine coupling with the electron spin, dipole-dipole or quadrupolar interactions is slow on the timescale of typical hops. Generally, the coupled electron-nuclear spin dynamics may by itself result in the non-exponential spin relaxation~\cite{PhysRevLett.97.037204,PhysRevB.84.155315,1367-2630-9-7-226,sin:quad}, in our case it results in the cut-off of the algebraic tail in the spin dynamics, see below. In the considered model, the electron spin at a given 
site $i$ rapidly precesses in the static hyperfine field $\bm \Omega_i$, as a result, only its 
projection onto $\bm \Omega_i$ is conserved during the typical hop 
waiting time~\cite{KT,merkulov02}. Then the electron hops to another site $j$ and its spin precesses in the field~$\bm \Omega_j$. Hence, the random hyperfine 
fields and hopping between the sites serve as a source of spin decoherence and relaxation. Specifics of spin decay in the opposite limit, $\Omega_0 \tau \ll 1$, due to the L\'{e}vy distribution of the waiting times were studied in Ref.~\cite{raikh16}.

It is important that the condition $na\ll\,1,$ ensuring the exponentially broad distribution of the hopping rates~\eqref{Wij},
leads to the situation where  for the most of the sites the electron hop to its 
nearest neighbor is exponentially more probable than all other hops. 
It allows us to assume that the electron always hops to its nearest neighbor site and divide all the sites into two groups. For the first group of ``pair'' sites the relation ``the nearest neighbor'' is mutual, i.e. the two ``pair'' sites are the nearest neighbors for one another, see Fig.~\ref{fig:model}, cf. with the cluster model of Ref.~\cite{gis2014noise}. The relaxation of the spin on ``pair'' sites is related to the multiple hops inside the pair. For the second group, the ``reaching'' sites, the ``nearest neighbor'' relation is not mutual. The ``reaching'' 
sites serve as the bottleneck for the spin relaxation: the longest waiting time in this situation is that of 
a hop for the site $i$ to its nearest neighbor $k$. 
The subsequent relaxation is due to the hops to other sites that are 
closer to $k$ than the site $i$ and occur much faster than the initial hop $i \rightarrow k$.
After several hops the electron reaches one of the ``pair'' sites and remains in this ``pair''. Note, that in our model the typical distances between the sites in ``pairs'' is large compared with the localization radius $a$ due to the condition~\eqref{kinetics}. The spin dynamics and noise in a two-site complex with the distance $\lesssim a$ where the quantum tunnelling between the sites is important has been studied in Ref.~\cite{pershin:qdm}.

\heading{Results.}  
The contributions of the two groups of states to the long-time spin dynamics can be evaluated separately. 
The relaxation of the spin initially located on 
a ``reaching'' site $i$ at times $t\gg\Omega_0^{-1}$ is determined by the rate of the fastest 
hop from this site to its neighbor. The expectation value 
of spin-$z$ component~\cite{note2} on the given ``reaching'' sites having the distance to the 
nearest neighbor close to $r_{i},$ can be estimated 
as $\langle s_{i,z}(t) /s_{i,z}(0) \rangle = s_R(t,r_i) = \exp(-t/\tau_i) K_{KT}(\Omega_0,t)$, 
where $\tau_i = \tau \exp(2r_i/a).$ Here 
$K_{KT} = \frac{1}{3}+\frac{2}{3}(1-\Omega_0^2t^2)\exp(-\Omega_0^2t^2/2)$ 
is the Kubo-Tayabe formula~\cite{merkulov02,KT}, which describes the spin precession in the 
static nuclear field $\bm \Omega_i$ resulting in 
the depolarization of transverse to $\bm \Omega_i$ spin components. Asymptotically 
(at $t \gg \Omega_0^{-1}$) $K_{KT}$ tends to $1/3$. 
The electron spin localized on sites $k$ and $l$ forming a ``pair'' relaxes according 
a similar exponential law $s_P(t, r_{kl}, \theta_{kl}) = \exp(-t/\tau_{kl}) K_{KT}(\Omega_0,t)$. Here $r_{kl}$ 
is the distance between the sites in the ``pair'' and $\theta_{kl}$ is the angle between the hyperfine fields at these sites, 
$\cos{\theta_{kl}} = (\bm\Omega_k \cdot \bm \Omega_l)/(\Omega_k \Omega_l)$ 
and time constant $\tau_{kl} = \tau \exp{(2r_{kl}/a)}/(1-|\cos \theta_{kl}|)$~\cite{PhysRevB.92.014206}. 
As a result, the disorder-averaged time dependence is given by
\begin{multline}\label{Smod}
\left\langle \frac{S_{{z}}(t)}{{S_z(0)}} \right\rangle 
= \int_0^{\infty} P_{R}(r_i) s_{R}(t,r_i) dr_i \\+ {\frac{1}{2}}\int_0^{\infty} dr_{kl} P_P(r_{kl}) 
\int_{-1}^{1} d(\cos\theta_{kl}) s_P(t, r_{kl}, \theta_{kl}).
\end{multline}
Here $P_{R}(r) = 2ne^{-2nr}(1-e^{-nr})$ is the probability for a site to be a ``reaching'' one 
with the distance to its neighbor equal to $r$ and $P_{P}(r)=2ne^{-3nr}$ is a probability 
for a site to be a part of a ``pair'' with the distance between sites $r$, and the Poisson 
distribution of sites is assumed.  Equation~\eqref{Smod} demonstrates that the total
spin dynamics is given by the weighted average of the electron spins on the ``reaching'' and ``pair'' sites. The integrals in Eq.~\eqref{Smod} can be expressed via special functions~\cite{suppl}.

\begin{figure}[t]
        \includegraphics[width=\linewidth]{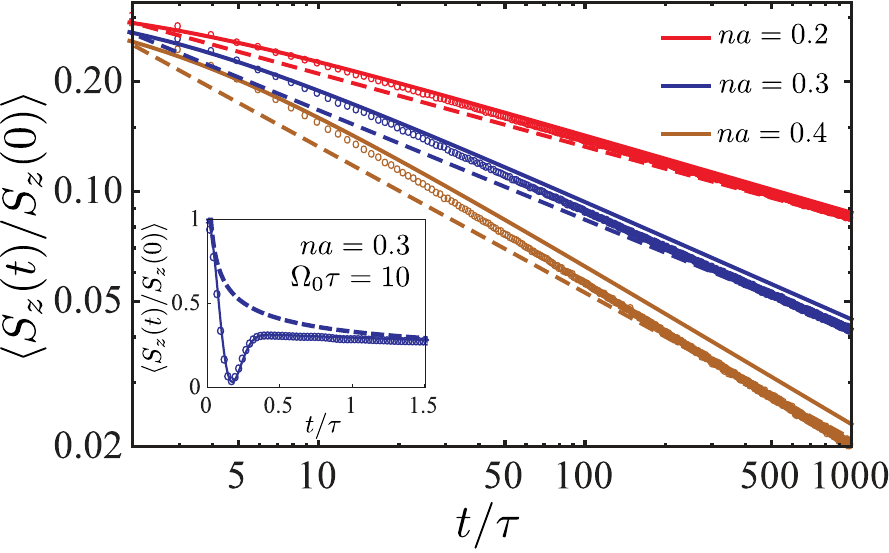}
        \caption{The comparison of the numerical results (dots) with the model of ``pair'' and ``reaching'' sites, Eq~\eqref{Smod}, (solid lines) 
        and power-law approximation Eq.~(\ref{pow1}) (dashed lines). 
        Inset shows the comparison on the initial time interval of the 
        relaxation for $na=0.3$ and $\Omega_0\tau=10$. Numerical calculation involved 100 sites and averaging over 5000 realizations of hyperfine fields. }
    \label{fig:timerel}
\end{figure}

To get insight into the spin dynamics of hopping electrons in nanowires, we performed numerical simulations based 
on the kinetic Eqs.~\eqref{set} and compared the results with the 
calculation after Eq.~\eqref{Smod}. Figure~\ref{fig:timerel} 
shows very good agreement of the model calculation (solid lines) with 
numerics (dots) for $na \lesssim 0.4$. Note that the  analytical model includes no free parameters. 

Now we turn to the detailed analysis of the long-time, $t \gg \tau$, spin dynamics. Figure~\ref{fig:timerel} indicates 
the power-law relaxation. To support this conjecture, we note that the spin at large $t \gg f\tau$, where 
the factor $f\gg 1$~\cite{note:f}, is governed by the 
sites with large distances to its neighbors, i.e. with large $r_i, r_{kl} \gg n^{-1}$ in Eq.~(\ref{Smod}). 
 The sites with large distance to its nearest neighbor are typically the ``reaching'' sites 
 because $P_R(r) \gg P_P(r)$ for $r \gg n^{-1}$. 
Indeed, let us consider 4 sites $(k,\dots,k+3)$ in a row. The sites $(k+1,k+2)$ form 
a pair if 
\begin{equation}
x_{k+1}-x_{k}>x_{k+2}-x_{k+1}<x_{k+3}-x_{k+2},
\end{equation}
while the sufficient condition for a site, e.g., $k+2$ to be a ``reaching'' one is
\begin{equation}
x_{k+1}-x_{k}<x_{k+2}-x_{k+1}<x_{k+3}-x_{k+2}.
\end{equation}
As a result, the distances in the ``pairs'' are
statistically smaller than the distances between the ``reaching'' sites and their nearest neighbors.
The ``reaching'' sites contribution into Eq.~\eqref{Smod} can be analytically 
evaluated at $na\ll 1$ and yields the $t\gg \tau$ asymptotics ~\cite{suppl}
\begin{equation}\label{pow1}
{\left\langle \frac{S_z(t)}{S_z(0)} \right\rangle = \frac{1}{3} }C(na) \left(\frac{t}{\tau}\right)^{-na},
\end{equation}
where a coefficient $C(na) \sim 1$. Equation~\eqref{pow1} clearly demonstrates that the 
long-time spin dynamics is described by the power law 
with the exponent controlled by the density of sites and the localization length. 
Since the ``reaching'' states serve as a bottleneck for the spin relaxation, the long-time 
spin evolution is controlled by the exponentially broad spread
of the waiting times for these sites. The numerical simulations show that the asymptotic (\ref{pow1}) 
with $C=1$ is in a good agreement with the numerical results in a wide range of times and parameters $na$~\cite{suppl}. 
Only at a small $t \lesssim \tau$ the asymptotics~\eqref{pow1} deviates from the numerical results,
that agree with the general model, Eq.~\eqref{Smod}, see the inset in Fig.~\ref{fig:timerel}.

\begin{figure}[ht]
        \includegraphics[width=\linewidth]{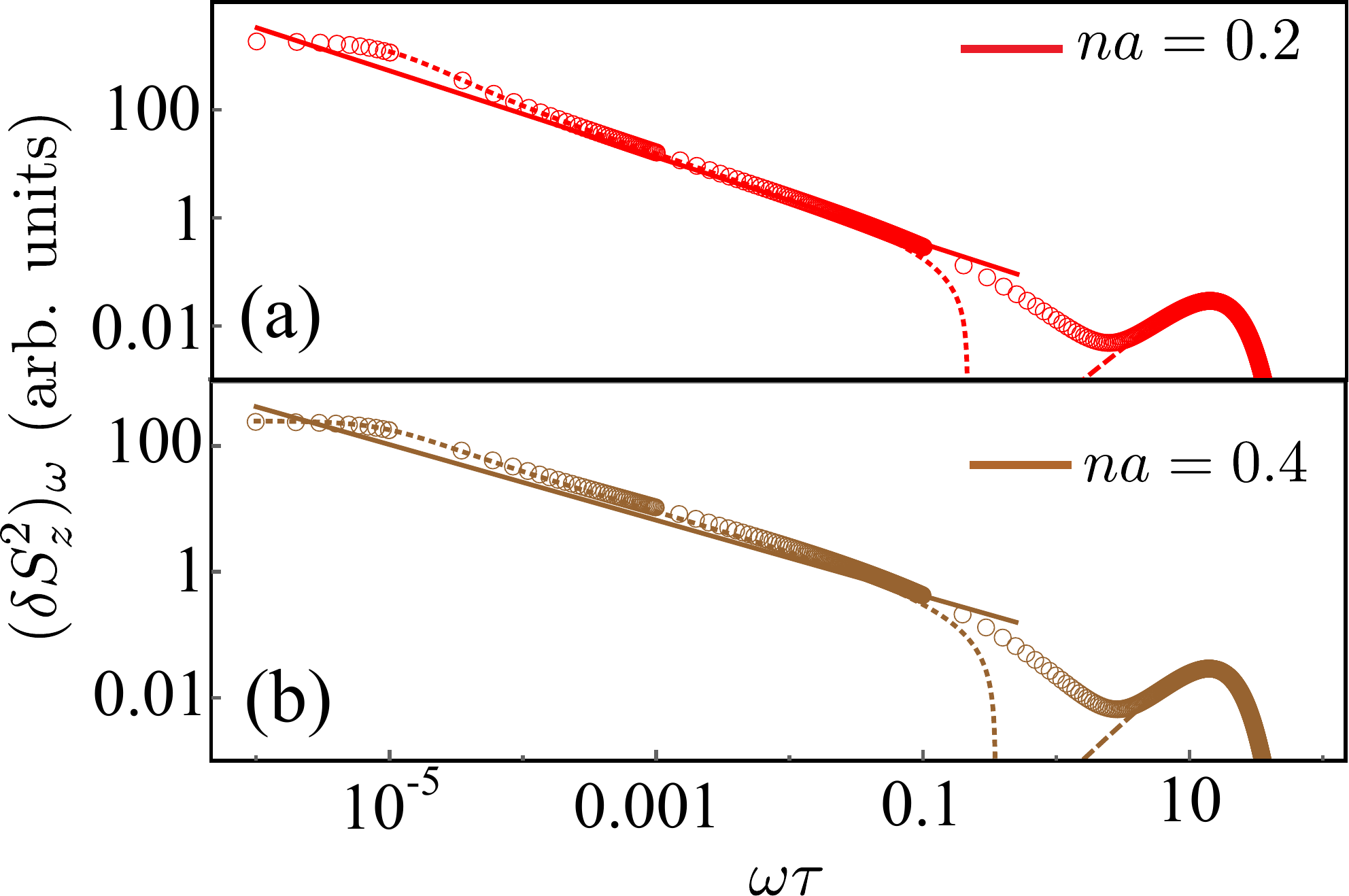}
        \caption{Spin noise spectra calculated using Eqs.~\eqref{Smod} and~\eqref{sns:def} (open circles), its power-law asymptotics, 
        Eq.~\eqref{SNS} (solid) and spin precession peak (dashed), $\Omega_0\tau=10$. 
        In the numerical calculation the exponential decay of 
        spin polarization $\propto \exp{(-t/\tau_s)}$ with $\tau_s/\tau=10^5$ was included. Dotted lines show asymptotics calculated with account for the spin relaxation time~\cite{suppl}.}
    \label{fig:sns}
\end{figure}

\heading{Spin noise.}
The power-law spin decay results in a zero-frequency anomaly in the spin noise spectrum. 
The autocorrelation function, $\langle \delta S_z(t) \delta S_z(t')\rangle$, 
obeys the same set of kinetic equations~\eqref{set}~\cite{PhysRevB.91.195301,2015arXiv151107661G}. 
Defining the spin noise power spectrum in a standard way, 
\begin{equation}
\label{sns:def}
(\delta S_z^2)_\omega = \int \langle \delta S_z(t) \delta S_z(t')\rangle \exp{(\mathrm i \omega t)} dt,
\end{equation} 
(see Refs.~\cite{gi2012noise,ivchenko73fluct_eng,2015arXiv151107661G,sin:rev} 
for details), we obtain by virtue of Eq.~\eqref{pow1} the power-law 
feature in the spin noise spectrum~\cite{suppl}:
\begin{equation}
\label{SNS}
(\delta S_z^2)_\omega = {\frac{\pi }{12} \frac{ \tau na C(na)}{(\omega\tau)^{1-na}}}, \quad \omega\tau \ll 1
\end{equation}
The $\omega^{na-1}$ feature in the spin noise spectrum is a direct consequence of the broad distribution of 
relaxation rates at the ``reaching'' sites. In general, the low-frequency spectrum is expected to 
be $1/\omega^{\gamma}.$ In the above presented model, $\gamma=1-na$ and we obtain the ``flicker'' noise 
in the low-concentration limit. The exact value of $\gamma$ depends on the system details \cite{suppl}.
It is noteworthy that the singularity in the spin noise 
spectrum at $\omega=0$ is smeared out due to nuclear spin dynamics: At the time scales on the order of 
the nuclear spin dephasing time, $T_{2N}$, caused, e.g., by the hyperfine coupling with the electron spin, the dipole-dipole coupling between neighboring 
nuclei or quadrupolar splittings, the frequencies $\bm \Omega_i$ in Eq.~\eqref{set} are no longer 
static resulting in the cut-off of the power-law feature in Eq.~\eqref{SNS} at $\omega \lesssim 1/T_{2N}$. 

Figure~\ref{fig:sns} shows the spin noise spectra calculated with Eqs.~\eqref{Smod} 
and \eqref{sns:def} in the model of ``pairs'' and ``reaching'' states for the two typical 
values of $na=0.2$ and $0.4$. In the numerical calculation we have also introduced the 
exponential cut-off of the spin polarization replacing $K_{KT}(\Omega_0,t)$ 
in the expression for the spins at the ``reaching'' and ``pair'' states by the 
product $K_{KT}(\Omega_0,t)\exp{(-t/\tau_s)}$, where $\tau_s$ is the 
phenomenological spin relaxation time related, e.g., with the nuclear spin dephasing. 
Three frequency ranges are clearly visible in Fig.~\ref{fig:sns}. At high 
frequencies $\omega  \sim \Omega_0$ ($\Omega_0\tau=10$ in our 
calculation) a peak in the spin noise spectrum is seen. It is related with 
the electron spin precession in the field of frozen nuclear fluctuation
and serves as direct evidence of the hyperfine coupling of the electron and 
nuclear spins~\cite{2015arXiv150605370B}. 
At this frequency range where $\omega \sim \Omega_0 \gg 1/(f\tau)$ the electron hopping 
is unimportant and the precession peak is described by the theory developed 
in Refs.~\cite{gi2012noise,PhysRevB.91.195301}, see dashed lines in Fig.~\ref{fig:sns}. 
In the range of intermediate frequencies $\tau_s^{-1} \ll  \omega \lesssim (f\tau)^{-1}$ 
the spin noise spectrum is well described by a power-law asymptotics, 
Eq.~\eqref{SNS}, see solid lines in Fig.~\ref{fig:sns}. This is a direct consequence of the algebraic 
tail in the spin relaxation, $S_z(t) \propto t^{-na}$, Eq.~\eqref{pow1}. 
Here, the power-law temporal decoherence of spins results in a power-law feature in the spin noise. 
Finally, at $\omega \lesssim \tau_s$ the algebraic 
singularity in the spin noise spectrum is suppressed and $(S_z^2)_\omega$ reaches a constant 
value $\propto \tau_s$ at $\omega\to 0$. 
Overall, the $\omega-$dependence of the spin noise power spectrum in the low-frequency domain is described 
by the dashed asymptotic presented in the Supplemental 
information~\cite{suppl}, which takes into account the exponential spin relaxation at $t\gtrsim \tau_s$.

\heading{Conclusion.} We have developed a theory of spin decoherence and corresponding spin 
noise in nanowires taking into account the key feature 
of disordered one-dimensional systems such as a strong electron localization resulting 
(i) in the electron hopping transport and (ii) enhanced 
hyperfine coupling between the electron and host lattice nuclear spins. 
As a result, the evolution of electron spin density fluctuations is
governed by the interplay of the electron hopping and hyperfine interaction. 
Here the localization, at least for a low density of sites, 
results in the exponentially broad distribution of the hop waiting times and, 
consequently, in the algebraic long-time spin decay
and in the power-law low-frequency singularity in the spin noise spectrum. It may be 
expected that the formation of the long-time spin decay due to exponentially wide distribution 
of waiting times is a general property of hopping electrons: In Ref.~\cite{PhysRevLett.110.176602} it was shown for the spin relaxation dominated by the spin-orbit coupling.

\heading{Acknowledgements.} A.V.S. acknowledges partial support from the RFBR project grant No.~15-02-01575. E.Y.S. acknowledges support of the University of the Basque Country UPV/EHU under program UFI 11/55, Spanish MEC (FIS2012-36673-C03-01 and
FIS2015-67161-P) and Grupos Consolidados UPV/EHU del Gobierno Vasco
(IT-472-10). M.M.G. was partially supported by RFBR project No.~14-02-00123, the Russian Federation President Grant MD-5726.2015.2, Dynasty foundation and SPbSU Grant No.~11.38.277.2014. We acknowledge  valuable communications with A. Pitanti and F. Rossella and D. Smirnov.


\begin{thebibliography}{24}
\bibitem{note2} Since the considered system is isotropic in the spin space, the same results 
are valid for the relaxation of any of the spin components.
\end{thebibliography}
\end{document}


\title{Supplemental Material for ``Spin dynamics of hopping electrons in quantum wires: algebraic decay and noise''}
\author{A.V. Shumilin}
\affiliation{Ioffe Institute, 194021 St.-Petersburg, Russia}

\author{E.Ya. Sherman}
\affiliation{Department of Physical Chemistry, The University of the Basque Country, 48080 Bilbao, Spain}
\affiliation{IKERBASQUE Basque Foundation for Science, Bilbao, Spain}

\author{M.M. Glazov}
\affiliation{Ioffe Institute, 194021 St.-Petersburg, Russia}
\affiliation{Spin Optics Laboratory, St.-Petersburg State University, 1 Ul'anovskaya, 198504 Peterhof, St.-Petersburg, Russia}
\maketitle

\section{Robustness to the distribution of the system parameters}\label{sec:rob}

Here we prove robustness of our approach to the spread of the system parameters and to the effects of strongly
temperature-dependent hopping probabilities. In addition, we will discuss the effects appearing due to a
finite concentration of electrons where the effects related with the occupancy of states becomes important.

\subsection{Robustness to the distribution of hopping matrix elements}

In the main text we assumed the hopping rate probabilities for the states
localized near the nanowire points $x_{i}$ and $x_{j}$ in the form, cf. Eq.~(3),
\begin{equation}
W_{ij}=\frac{1}{\tau}
\exp\left( -2\frac{\left|x_{i}-x_{j}\right|}{a}\right),
\label{distribution}
\end{equation}%
were $\tau $ is determined by the electron-phonon coupling and $a$ is the
characteristic spread of the wavefunction along the wire axis. Here we
estimate the typical for III-V semiconductors values of $\tau$ and show
that the results in the main text are robust against the realistic spread of
$\tau$ and $a$ determined by a typical disorder in the system.

Localized wavefunctions $\psi_{i}\left( \mathbf{r}\right)$
(with the corresponding eigenenergies $E_{i}$) are orthogonal
to each other normalized eigenfunctions of the full Hamiltonian including the
effect of the disorder. A typical spread of these energies is of the order of few meV.
Electron hopping between the states with the energies $E_{i}$ and $E_{j}$ is
accompanied by absorption or emission of a phonon with the momentum $%
q=\left\vert E_{i}-E_{j}\right\vert /\hbar c$ and occupation factor (at
temperature $T\gg \left\vert E_{i}-E_{j}\right\vert $) of $%
n_{q}=T/\left\vert E_{i}-E_{j}\right\vert $. This momentum determines the
electron-phonon coupling strength, the phonon density of states, and the
wave-function dependent transition matrix elements given by the Fourier
transform
\begin{equation}
M_{ij}\left( \mathbf{q}\right) =\int \psi_{i}^{\ast }\left( \mathbf{r}\right)
\exp \left(\pm \mathrm i \mathbf{q}\cdot\mathbf{r}\right) \psi_{j}\left(\mathbf{r}\right) d\mathbf r.
\end{equation}%
In the $E_{i}-E_{j}=0$ limit  the matrix element $M_{ij}\left( \mathbf{q}%
\right) $ vanishes due to orthogonality of the wave functions, and in the
limit of large $q$ it is small due to the fact that asymptotic of the
Fourier transform of a regular function should vanish at a large momentum. The
large-${q}$ asymptotic strongly depends on the details of the shape of
the wave functions. Accordingly, we model the hopping satisfying these
conditions with the expression
\begin{multline}
W_{ij}=\frac{1}{\tau_{D}}
\frac{\Delta ^{3}\left| E_{i}-E_{j}\right| T}{\Delta ^{5}+\left| E_{i}-E_{j}\right| ^{5}} \times 
\\
\left[\exp\left({-2\frac{\left| x_{i}-x_{j}\right|}{a_{i}}}\right)+
\exp\left({-2\frac{\left| x_{i}-x_{j}\right|}{a_{j}}}\right)
\right],
\label{distribution2}
\end{multline}%
vanishing at small and large $\left|E_{i}-E_{j}\right|.$
Here $a_{i}$ and $a_{j}$ are the localization lengths for corresponding sites,
and the parameter $\Delta\equiv c\hbar/a,$ where $a$ is the typical localization length and
$c$ is the longitudinal sound speed, determines the crossover between small and large momentum regimes, with the crossover
momentum $q=1/a.$ For qualitative analysis we assume that electro-phonon
coupling is due to the deformation potential, neglecting a relatively small contribution
of piezointeraction \cite{Gantmakher,Beya}. According to the
Fermi's golden rule, the characteristic hopping time
determined by the electron-phonon coupling is estimated as $\tau^{-1}_{D}={D^{2}}/{\zeta}a^{3}{\hbar c^{2}}.$
The values $\Delta $ and $\tau _{D}$ strongly depend on the system
parameters. For the density $\zeta
=5$ g/cm$^{3},$  $c=5\times 10^{5}$ cm/s, and $a=10$ nm
we obtain $\Delta \approx 0.3$ meV. The typical
deformation potential $D=7.5$ eV \cite{Priester,Cardona} yields $\tau _{D}\approx 10^{-2}$ ns.
The hyperfine-induced spin precession rate $\Omega_{0}$ is of the order of $\sim 10^{9}$ s$^{-1}.$
Bearing in mind that for $na<0.4$, the exponent $\exp(2/na)$ is larger than $10^{2}$, this
estimate implies that  between the intersite jumps the electron spin experiences many rotations
in the hyperfine field. It is worth mentioning that
the slow algebraic decay of spin is observed already at times much smaller than $\tau\exp(2/na)$
(see footnote [36] in the main text and Sec.~\ref{sec:asympt}). Recalling that typical times for nuclear motion are of the order of
or exceed $\sim\,10^{3}\Omega_{0}^{-1}$~\cite{Merkulov}, we  confirm the existence of the time interval where
the electron hopping with long waiting times occurs while the nuclear hyperfine fields are still static.

To prove the robustness of our results we model the spin relaxation with the
distribution of  probabilities (\ref{distribution2}) where the random
energies are uniformly distributed in the interval $(E_{0}-\Delta E/2, E_{0}+\Delta E/2)$.
To evaluate the corresponding wavefunction spread, we use the semiclassical asymptotic with $a_{i}=a_{0}\sqrt{E_{0}/E_{i}},$
where $a_{0}$ corresponds to the $E_{0}$ energy.

\begin{figure}
    \centering
        \includegraphics[width=\linewidth]{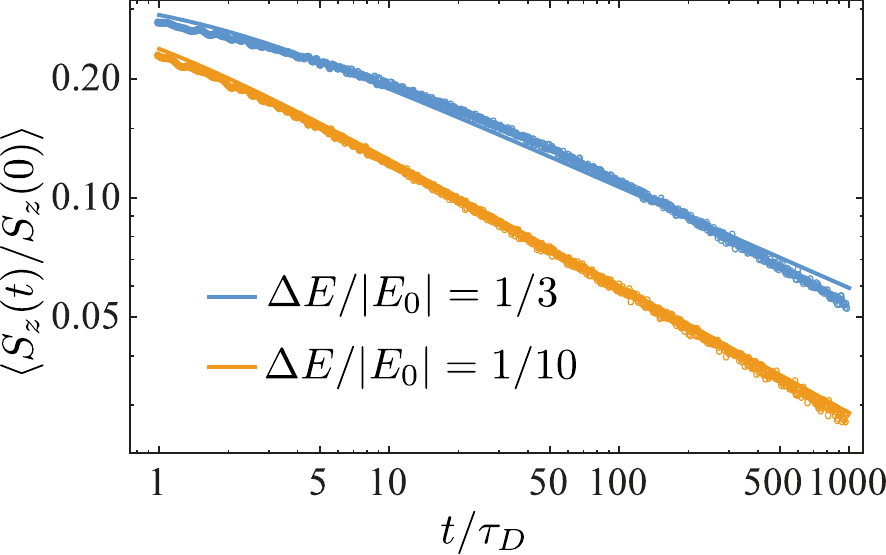}
        \caption{The spin relaxation with finite temperature and random on-site energies. Points correspond to the numerical simulation
        based on Eq.~(2) of the main text, lines correspond to the analytical model, Eq.~(4) of the main text, with the parameters $a$ and $\tau$ fitted to reproduce the results of numerical simulation.}
    \label{fig:robust}
\end{figure}

The temperature and the energy cutoffs
are taken as $T=\left|E_{0}\right|/3$ and $\Delta = 0.05\left|E_{0}\right|$. The results for $\Delta E=\left|E_{0}\right|{/10}$
and $\Delta E=\left|E_{0}\right|/3$ are shown in Fig.~\ref{fig:robust} and compared with
Eq. (4) of the main text with the $\Delta E$-dependent effective parameters $\tau$ and $a$.
The parameters used in this comparison are $\tau = 0.25 \tau_{D}$, $a=a_{0}$ for $\Delta E = 0.1 \left|E_{0}\right|$
and $\tau=0.6\tau_D$, $a=0.8a_0$ for $\Delta E =\left|E_{0}\right|/3$. The results show that although
the distribution of the pre-exponential
factors in the hopping probabilities and energy-dependent localization length lead
to some modification of the model parameters, the main features of the relaxation remain unchanged,
proving the robustness of our model.

\subsection{Effects of temperature and the electron concentration.}

In the main text we focused on the system realization with a small mean occupation of localized states
and the temperature considerably higher than their
random energy spread, that is $T\gg |E_j-E_i|.$  Here we discuss how the moderately low temperatures and finite electron concentrations modify the proposed picture.

The kinetic equations similar to Eq. (2) of the main text for finite electron concentrations read \cite{PhysRevB.92.014206}
\begin{multline}\label{kin-T}
\frac{d \bm S_i}{dt} + \bm S_i \times \bm \Omega_i =\\ 
\sum_j \left[W_{ji} (1-N_i) \bm S_j - W_{ij}(1-N_i) \bm S_i\right],
\end{multline}
where 
$$N_i = \frac{1}{1+\frac{1}{2}\exp{\left(\frac{E_i-{\mu}}{T}\right)}}$$ 
is the mean filling of the site $j$ and ${\mu}$ is the chemical potential. Doubly occupied states are neglected.
The hopping probabilities contain
the temperature-dependent exponent as:
\begin{multline}\label{Wij-T}
W_{ij} = \frac{1}{\tau_0} \exp\left(- \frac{|x_j-x_i|}{2a}  \right) \times 
\\
\left[ \theta(E_i-E_j) +  \exp\left(- \frac{|E_j-E_i|}{T} \right) \right].
\end{multline}
Here we consider the characteristic energy differences $\left|E_j-E_i\right|$ to be larger than the temperature.
We also note that the derivation of (\ref{kin-T}) disregards the correlations in the
on-site density matrix and spin-spin exchange
interaction - both effects can be important for sufficiently large occupations.
Nevertheless, the Eqs.  (\ref{kin-T}) and (\ref{Wij-T})
can be used to qualitatively understand the suppression of the discussed
relaxation and noise with decreasing temperature.

\begin{figure}
    \centering
        \includegraphics[width=\linewidth]{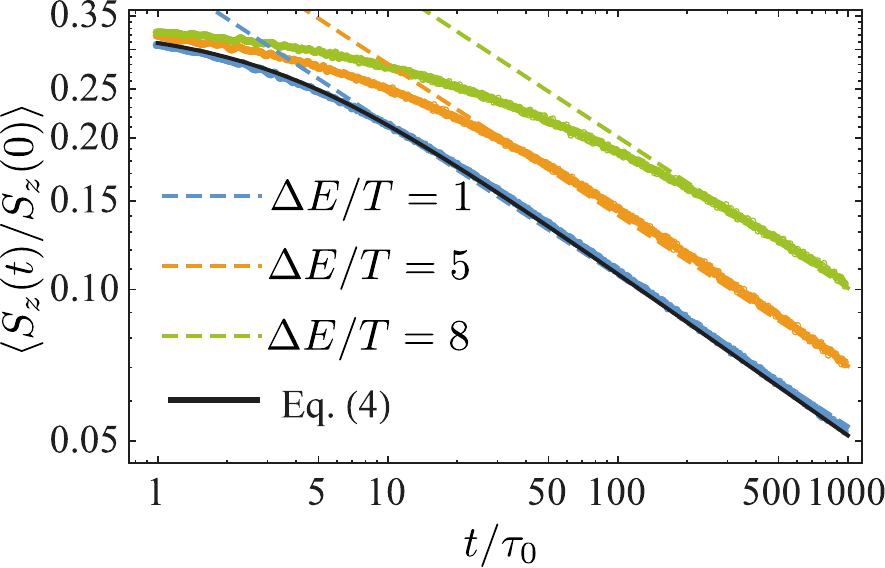}
        \caption{The spin relaxation with finite temperature and random on-site energies.
        Points correspond to the numerical simulation
        based on eq.~(\ref{kin-T}), dashed lines correspond to the asymptotic $A/t^{na}$ and
        the black line corresponds to the Eq.~(4) from the main text.}
    \label{fig:pict-T}
\end{figure}

Figure \ref{fig:pict-T} shows the numerical simulation based on Eq\addMisha{s}. (\ref{kin-T}) and \eqref{Wij-T}. The site energies were randomly chosen in the
interval $({\mu}-0.1\Delta E, {\mu}+0.9\Delta E )$ near the chemical potential level. It
ensures that the electron concentration $n_e$ is relatively small compared to the site concentration $n$: $n_e/n\approx 0.18$ for  the $\Delta E/T=8$ and $n_e/n\approx0.5$ for the $\Delta E/T=1$. The spatial distribution of the sites is Poissonian with $na=0.3$. The results are compared with the results of Eq.~(4)
from the main text and with the power-law asymptotic $A/t^{na}$.

For relatively small $\Delta E \sim T$ the calculations reproduce very well the relaxation considered in the
main text. They agree with Eq.~(4) from the main text with parameter $\tau=1.5\tau_0$. However when $\Delta E$ is large the
initial part of relaxation at times $t \sim \tau$ is suppressed. This relaxation is
related to the pairs of sites that are separated by a
distance $r \lesssim a$ and both have the energy near the Fermi level. Such pairs
become rare when $\Delta E \gg T$. At sufficiently
large times the relaxation behaves as $S_{z}(t)=A/t^{na}$ with the power $na$ insensitive to~$\Delta E$.

The independence of the power law asymptotic of the random energies is related to the sharp cutoff
of the energies at the interval boundaries. When the exponential band tails are included
into consideration, the algebraic relaxation with a temperature-dependent power appears even
without a spatial disorder \cite{PhysRevLett.110.176602}.

\section{Asymptotes  of spin decay and of spin noise}\label{sec:asympt}

Equation (4) of the main text can be presented in the form:
\begin{equation}\label{Smod}
\left\langle \frac{S_{{z}}(t)}{{S_z(0)}} \right\rangle = \left\langle \frac{S_{{z}}(t)}{{S_z(0)}} \right \rangle_r + \left\langle \frac{S_{{z}}(t)}{{S_z(0)}} \right\rangle_p, 
\end{equation}
with
\[
 \left\langle \frac{S_{{z}}(t)}{{S_z(0)}} \right \rangle_r = \int_0^{\infty} P_{R}(r_i) s_{R}(t,r_i) dr_i,
\]
and
\[
 \left\langle \frac{S_{{z}}(t)}{{S_z(0)}} \right\rangle_p = {\frac{1}{2}}\int_0^{\infty} dr_{kl} P_P(r_{kl}) \int_{-1}^{1} d(\cos\theta_{kl}) s_P(t, r_{kl}, \theta_{kl}).
\]
These integrals can be evaluated analytically at $t\gg 1/\Omega_0$ via special functions. For the reaching states contribution one has:
\begin{widetext}
\begin{equation}
\label{reaching}
\left\langle \frac{S_{{z}}(t)}{{S_z(0)}} \right \rangle_r = \frac{na}{3} \left(\frac{t}{\tau}\right)^{-3 na/2}  
\left\{\left(\frac{t}{\tau}\right)^{{na}/{2}} [\Gamma (na)-\Gamma (na,t/\tau)]+\Gamma \left(\frac{3
   na}{2},\frac{t}{\tau}\right)-\Gamma \left(\frac{3 na}{2}\right)\right\},
\end{equation}
where $\Gamma(x)$ is the $\Gamma$-function and $\Gamma(x,y)$ is the incomplete $\Gamma$-function defined as  $$\Gamma(x,y) = \int_y^\infty u^{x-1} e^{-u} du.$$ 
The contribution of pairs reads
\begin{equation}
\label{pair}
\left\langle \frac{S_{{z}}(t)}{{S_z(0)}} \right \rangle_p = \frac{na\tau}{3t} \left[{\rm E}_{\nu}(t/\tau)+\frac{2}{3
   {na}-2}\right]
-\frac{na}{3\tau} t^{-3 {na}/2} \Gamma \left(\frac{3
   {na}}{2}-1\right),
\end{equation}
\end{widetext}
where 
\[
\nu = 2-\frac{3 {na}}{2},
\]
and the generalized exponential integral function 
\[
{\rm E}_{\nu}(x) = \int_1^\infty \frac{e^{-xu}}{u^\nu} du = x^{\nu -1} \Gamma(1-\nu,x),
\]
are introduced.

\begin{figure}[h!]
    \centering
        \includegraphics[width=\linewidth]{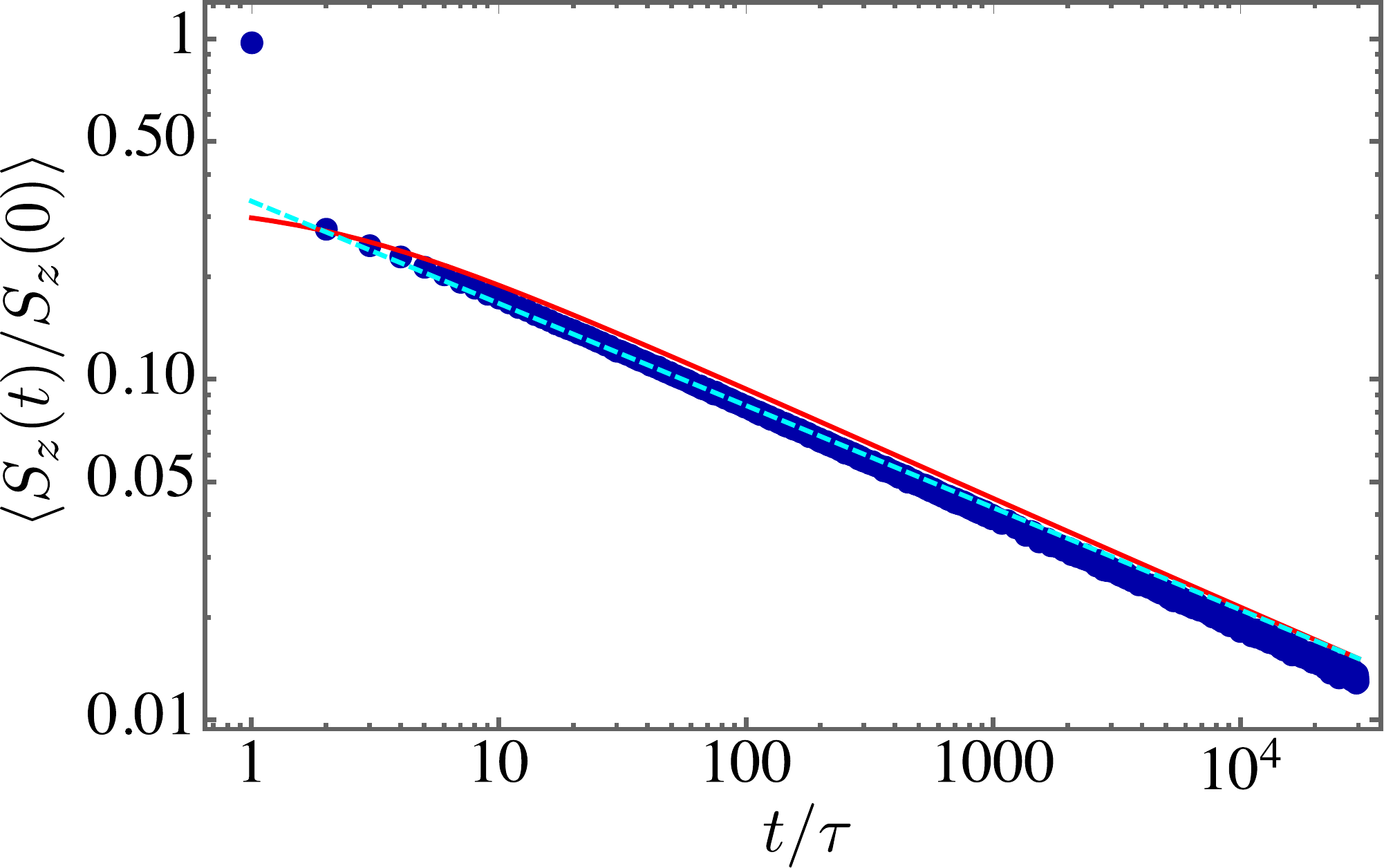}
        \caption{Numerical simulation of the long-time behavior of the spin decay (circles) for $na=0.3$ calculated in the model of the main text. Dashed cyan line shows the asymptotics presented in Eq.~(7) of the main text with $C=1$ and red line shows the analytical result, Eqs.~\eqref{Smod}, \eqref{reaching}, and \eqref{pair}. We used 100 sites and averaged the result over 1700 realizations.}
    \label{fig:long}
\end{figure}

Linear-$t$ asymptotics (valid at the short times, formally $1/\Omega_0 \ll t \ll \tau$):
\begin{equation}
\label{asympt:short}
\left\langle\frac{S_z(t)}{S_z(0)}\right \rangle = \frac{1}{3} - \frac{na (1+2na)t}{3(1+na)(2+3na)}.
\end{equation}
At long times, $t\gg \tau \exp{(2/na)}$ we have
\begin{equation}
\label{asympt:long}
\left\langle\frac{S_z(t)}{S_z(0)}\right \rangle = \frac{\Gamma(na)}{3} \left(\frac{t}{\tau}\right)^{-na}.
\end{equation}
This leading order asymptotics results from the reaching states only, Eq.~\eqref{reaching}.
At $na\to 0$ the $\Gamma(na)\to 1$ and Eq.~\eqref{asympt:long} is in agreement with Eq.~(7) of the main text. Figure~\ref{fig:long} shows the comparison of the numerical simulation (points) with the analytical formulae Eqs.~\eqref{Smod}, \eqref{reaching}, and \eqref{pair} (red line) and asymptotics~(7) of the main text. Overall, a good agreement is observed.

Spin noise spectrum is defined by [cf. Eq. (8) of the main text]
\begin{equation}
\label{sns:gen}
(S_z^2)_\omega = \int \langle S_z(t+t')S_z(t)\rangle \exp{(\mathrm i \omega t')} dt' 
\end{equation}
\[
= \frac{1}{4} \int \left\langle  \frac{S_z(t')}{S_z(0)}\right \rangle  \exp{(\mathrm i \omega t')} dt',
\]
where the factor $1/4$ in the second equality takes into account the fact that the single-time correlator of spin components equals to $1/4=(1/3)\cdot 1/2 \cdot (1/2+1)$. Taking $\langle S_z(t)/S_z(0) \rangle$  in the form of Eq. (7) of the main text multiplied by $\exp{(-t/\tau_s)}$ where $\tau_s$ is the phenomenological spin relaxation time (cut-off time, related, e.g, with nuclear spin dynamics) we get in the limit of $\omega \ll 1/\tau$:
\begin{multline}
\label{sns:low}
(S_z^2)_\omega = \frac{C(na)}{6} \Re \left\{{\rm E}_{na}\left(\frac{\tau}{\tau_s} - \mathrm i \omega\tau \right) \right\} \\
\approx \frac{C(na)}{6} \left(\frac{\tau}{\tau_s}\right)^{na-1} \Re\{(1-\mathrm i \omega\tau_s)^{na-1} \} .
\end{multline}
This expression passes to Eq. (9) of the main text at $\tau_s \to \infty$.